\documentclass[letter]{aa}

\usepackage{graphicx}
\usepackage{epsfig}     % for figure inclusion
\usepackage{amsmath}
\usepackage{txfonts}
\usepackage{natbib}
\usepackage{multirow}
\bibpunct{(}{)}{;}{a}{}{,}
\usepackage{url}
\begin{document}
  \title{The large scale dust lanes of the Galactic bar}
  \author{D.J. Marshall\inst{1,3}
    \and
    R.Fux\inst{2}\fnmsep\thanks{Free-time astronomer}
    \and
    A.C. Robin
    \inst{1}
    \and
    C. Reyl\'e
    \inst{1}
  }
  \authorrunning{Marshall et al.}
  \institute{CNRS-UMR 6213, Institut Utinam, Observatoire de Besan\c{c}on, 
    BP 1615, 25010 Besan\c{c}on Cedex, France
    \and
    Observatoire de Gen\`eve, Universit\'e de Gen\`eve, 51 Ch. des Maillettes, 1290 Sauverny, Switzerland
    \and
    D\'epartement de Physique et Centre Observatoire du Mont M\'egantic, Universit\'e Laval, Qu\'ebec, QC, G1K 7P4, Canada  }
\date{Received 31 October 2007 / Accepted 14 November 2007 }

\newcommand{\changes}{}

\abstract
%CONTEXT
{Trails of dust inside galactic bars are easily observable in external galaxies. However, information 
on the dust lanes of the Milky Way bar is harder to obtain due to our position within the Galactic disc.}
%AIMS
{By comparing the distribution of dust and gas in the central regions of the Galaxy, we aim
 to obtain new insights into the properties of the offset dust lanes 
leading the bar's major axis in the Milky Way.}
%METHOD
{On the one hand, the molecular emission of the dust lanes is extracted from the observed 
CO $l-b-V$ distribution according to the interpretation of a dynamical model.
On the other hand, a three
dimensional extinction map of the Galactic central region constructed from near-infrared 
observations is used as a tracer of the dust itself and clearly reveals
 dust lanes in its face-on projection. Comparison of the position of both independent 
detections of the dust lanes is performed in the $(l,b)$ plane. }
%RESULTS
{These two 
completely independent methods are used to  
 provide a coherent picture of the dust lanes in the Milky Way bar. 
In both the gas and dust distributions, the dust lanes are found to 
be out of the Galactic plane, appearing at negative latitudes for $l>0\degr$ 
and at positive latitudes for $l<0\degr$. 
However, even though there is substantial overlap between the two components, they are offset from one another with the dust appearing to 
lie closer to the $b=0\degr$ plane.
}
%CONCLUSIONS
{
Two scenarios are proposed to explain the observed offset. The first involves grain destruction by the bar shock and reformation downstream. Due 
to the decrease in velocity caused by the shock, this occurs at lower $z$. 
The second assumes that the gas and dust remain on a common tilted plane,  but that the 
molecular gas decouples from the Milky Way's magnetic field, itself strong enough to resist the shear of the bar's shock. 
The diffuse gas and dust remain coupled to the field and are carried further downstream.
 This second scenario has recently 
been suggested in order to explain observations of the barred galaxy NGC 1097.
}


\maketitle

\section{Introduction}

Barred galaxies are ubiquitous in the Universe. Many of these are observed to have 
dust lanes - trails of gas and dust, seen in extinction, leading the bar's major axis.
Evidence for the bar nature of the Milky Way's bulge has been 
around for over 40 years \citep{deVaucouleurs1964}. However, our position in our Galaxy does not 
give us a clear view of the inner regions. 
Photometric studies are hampered by the high source confusion and 
severe interstellar extinction whereas gas kinematic studies are 
rendered difficult due to line of sight crowding of the gas emission lines.

Nevertheless, many studies have obtained information on the Milky Way's dust lanes using 
a variety of techniques. 
{\changes \cite{Calbet1996} detected a longitude assymetry at $10\degr<|l|<20\degr$ in the DIRBE surface brightness maps 
which they attributed to the dust lanes.} 
\cite{Babusiaux2005} used red clump stars to determine the distance 
as a function of longitude to
 the bar and using their observations concluded that there is no near-infrared extinction 
due to the dust lanes in our Galaxy. 
Other studies have based their interpretation of the observations on models in order to 
study the chemistry in the Galactic centre (GC) and dust lanes \citep{Rodriguez-Fernandez2006, Nagai2007}
as well as the physical processes underway in the dust lane shocks \citep{Liszt2006}. 
However, the results may depend on the assumed geometry for the dust lanes as no 
coherent picture currently exists for them.

In this letter, we combine the results from gas dynamics and stellar reddening 
studies in order to provide a description of the dust lanes leading the bar's major axis. 
\cite{Fux1999} interpreted the HI and CO longitude-velocity diagrams of the 
Galactic disc using self-consistent N-body and smooth particle hydrodynamics 
 simulations of the stellar and gaseous components. 
Later, \cite{Marshall2006} used a galactic stellar population synthesis model to 
extract the three dimensional extinction information from near infrared stellar observations. 
Both studies were able to identify the Milky Way's dust lanes.
In the following, 
we present and analyse the dusty and gassy sky distributions of the dust lanes resulting from these studies.

\section{Dust lanes from 3D extinction maps}\label{sec:marshall}

\cite{Marshall2006} created a three dimensional map of the inner Galaxy 
using the Two Micron All Sky Survey (2MASS) 
point source catalogue (PSC) \citep{Skrutskie2006} and the Stellar Population Synthesis Model, 
developed in Besan\c{c}on \citep[][which we will simply call ``Galactic model'']{Robin2003}. 
For a full description of the 3D extinction method, please see \cite{Marshall2006}.

Contrary to \cite{Marshall2006}, the method used in this study does not have a fixed angular 
resolution but one that varies as a function of the stellar density. The extinction information for the 
GC region presented here has an angular resolution that varies from $3.75 \arcmin$ (high density) 
to $15 \arcmin$ (low density).

The Galactic model used here differs slightly from that described in \cite{Robin2003}, 
as a result of more recent adjustments using 2MASS data.
Here, the stellar density of the bulge is modelled using the ${\rm sech}^2$ function from \cite{Freudenreich1998}. 
It is boxy and prolate, with axis ratios 1 : 0.3 : 0.25, and it is assumed to 
form an angle of 20$\degr$ with the Sun - GC direction, in line with recent estimates \citep{Gerhard2002}. 

\begin{figure}
  \centering
    \includegraphics[trim = 6mm 4mm 0mm 0mm ,clip, width=0.75\linewidth]{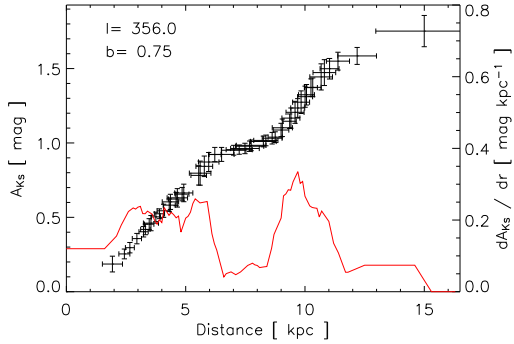}
  \caption{Application of the extinction method to the line of sight $l=356\degr$, $b=0.75\degr$.
    The crosses represent the total extinction as a function of distance (left hand scale) and the solid 
    red line is the corresponding differential extinction along the same line of sight (right hand scale). 
    The differential extinction is assumed proportional to the dust density.}
  \label{fig:example}
\end{figure}

The three dimensional extinction distribution for an example line of sight is shown 
in Figure \ref{fig:example}. Each cross represents an estimation of the total extinction at the given distance.
In order to visualise the three dimensional dust distribution we will present a map 
of the differential extinction, which is the derivative of the extinction with respect to distance. 
This will be proportional to the density of the big dust grains 
($15 {\rm nm} < $ radius $ < 110 {\rm nm}$) as they dominate the extinction  at 2 $\mu$m \citep{Desert1990}.
The differential extinction for the example line of sight is shown as the solid red line 
in Figure \ref{fig:example}. Under $\sim2$ kpc and
 at very large distances, where the distance-extinction points 
are spaced far apart, it is difficult to identify any underlying structure. 
However, the technique is well suited to studies of the Galactic bar : 
the extinction due to the dust lanes for this line of sight can be seen at $r\sim$9.5 kpc.
 
\begin{figure}
  \begin{center}
     \leavevmode
    \includegraphics[trim = 3mm 1mm 0mm 8mm ,clip, width=0.75\linewidth]{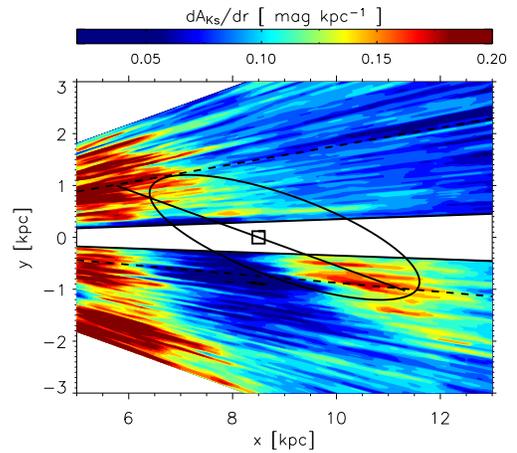}
  \end{center}
  \caption{The average density of absorbing matter in the 
    Galactic bar region for $|z|<300$ pc, as seen from the NGP. The quantity plotted 
is the differential extinction (in units of magnitudes per kpc in the $K_s$ band). 
The GC is indicated by the 
square symbol, the solid line shows the orientation of the modelled stellar bar, and the ellipse 
shows the region  selected{\changes, by inspection,} to contain the extinction arising from the dust lanes. The Sun is 
at (0,0), outside the figure. The dashed lines 
indicate galactic longitudes $l=-5\degr$ and $l=10\degr$ which approximately
 delimit the ends of the dust lanes as seen in CO.  
The white band indicates where the extinction estimates do not 
fully sample the bar due to the low completeness (high source confusion) of the 2MASS PSC.}
  \label{fig:ngp_dustlanes}
\end{figure}

The distribution of dust in the Galactic plane, towards the GC, 
is displayed in Figure \ref{fig:ngp_dustlanes}.
The elongated structure seen in 
extinction, running along the stellar bar (shown by the solid line in Fig.\ref{fig:ngp_dustlanes}) 
but with a slightly different angle, is due to the dust lanes. They can 
be seen to precede the bar in its clockwise motion (as seen from the NGP), 
in agreement with observations of dust lanes in external galaxies.
The near-side dust lane 
becomes intertwined with the molecular ring near $x=6$ kpc, $y=1$ kpc and the far-side dust lane reaches out 
to distances where the extinction map becomes unreliable or non-existent. 

\section{Dust lanes from molecular gas kinematics}\label{sec:fux}

\begin{figure}
    \centering
   \includegraphics[width=0.8\linewidth]{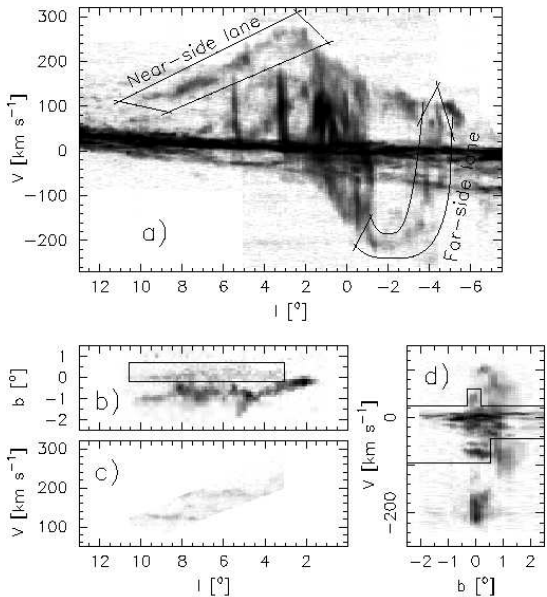}
    \caption{Selection procedure of the $^{12}$CO emission associated
    with the Galactic bar dust lanes, based on the
 \cite{Dame2001} $l$-$b$-$V$ data cube.
    {\bf a)}~Longitude-velocity diagram of the CO gas within $|b|\leq
    2.5^{\circ}$, with the black contours indicating the near-side
    (Connecting arm) and far-side lane regions set by our first
    selection criterion.
    {\bf b)}~Longitude-latitude diagram of the gas within the near-side
    lane region.
    {\bf c)}~Longitude-velocity diagram of the gas enclosed by the
    black box in the former $l$-$b$ plot; this contribution is
    rejected according to our second selection criterion.
    {\bf d)}~Longitude-latitude diagram of the gas in the far-side lane
    $l$-$V$ region; our last criterion excludes the emission between
    the two displayed lines. 
}
    \label{A}
 \end{figure}

Figure~\ref{A}a shows the observed $l-V$ distribution of $^{12}$CO
in the Galactic bar region. The black contours enclose the two
features which have been identified by \cite{Fux1999} as the gaseous
signature of the dust lanes along the Galactic bar. The $l>0\degr$
feature, also called ``Connecting arm'', is the trace of the dust lane
in the near-side part of the bar, and the $l<0\degr$ feature is the
trace of the far-side dust lane.
\par Because of the distance indetermination inherent to $l-b-V$
observations, one cannot entirely isolate the emission of these
dust lanes from foreground and background emission. The following
describes our approach to select most of the dust lanes CO emission
while excluding most of the unrelated emission. The first of the
adopted selection criteria restricts the emission to the $l-V$
regions highlighted in Figure~\ref{A}a. This criterion avoids in
particular the emission of the $|l|\la 1.5\degr$ nuclear
ring/disc and of Bania's Clump 1 \citep{Bania1977} at
$(l,V)\approx(-5\degr,100\,{\rm km\,s}^{-1})$. It also avoids
the emission of the velocity elongated features near $l=3\degr$
(Bania's Clump 2) and $l=5.5\degr$, which may arise from molecular
clouds being absorbed by the near-side dust lane 
\citep{Fux1999,Liszt2006,Rodriguez-Fernandez2006}. 

The
second criterion concerns the gas so far selected in the near-side
lane $l-V$ region. The $l-b$ distribution of this gas (Fig.~\ref{A}b)
reveals independent emission within
$3.125\degr\leq l\leq 10.5\degr$ and
$-0.125\degr\leq b\leq 0.625\degr$ (black box in the figure)
that forms two distinct $l-V$ features (Fig.~\ref{A}c). The nearly
constant velocity orientation of these features contrasts with the
global $l-V$ inclination of the near-side lane and thus suggests
that they do not belong to this lane. The gas within the box in
Fig.~\ref{A}b is therefore discarded. Finally, the third selection
criterion concerns the gas selected in the far-side lane $l-V$
region. The $b-V$ distribution of this gas (Fig.~\ref{A}d) separates
fairly well the emission of the dust lane from the bright emission of
external spiral arms, near $V=0\, {\rm km\, s^{-1}}$ and centred at $b\approx 0\degr$, and of
the 3-kpc arm, at $(b,V)\approx(0\degr,-75\,{\rm km\,s}^{-1})$. Hence we
also exclude all emission between the two lines drawn in
Fig.~{\ref{A}d}.
It should be noted that the remaining
emission at $l\ga 8.5\degr$ has characteristics similar to those
of the features seen in Fig.~\ref{A}c and thus may also not originate
from the dust lanes.

\section{Results and discussion}\label{sec:results}

\begin{figure}
   \centering
    \includegraphics[trim = 5mm 2mm 2mm 8mm ,clip, width=0.9\linewidth]{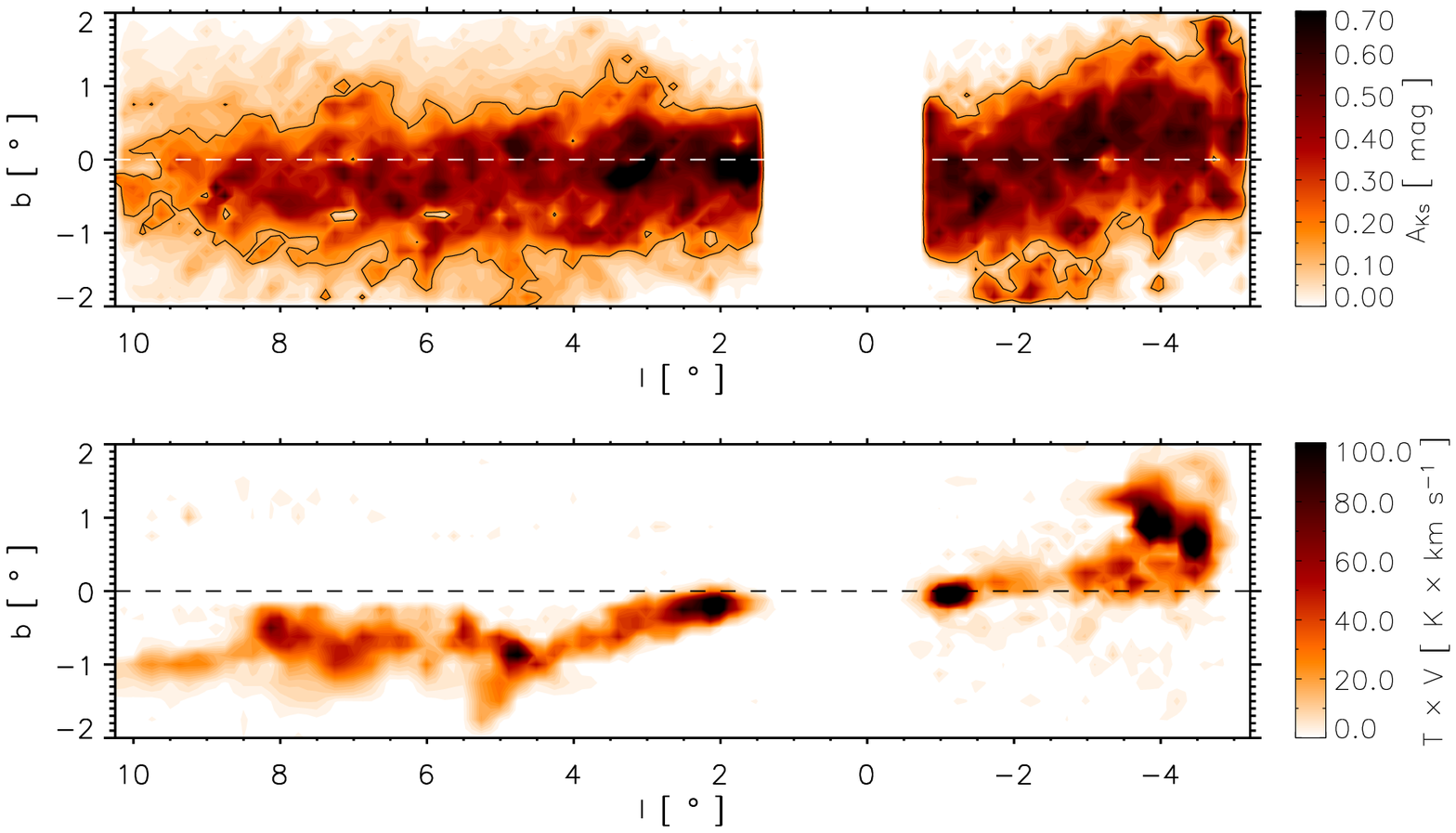}
   \caption{Longitude-latitude distributions of the near-infrared
   absorbing dust lanes detected in the Galactic bar by 
\cite{Marshall2006} ({\bf top}), and of the corresponding $^{12}$CO
   emission derived from the \cite{Dame2001} data assuming
   the $l$-$V$ feature interpretation given by \cite{Fux1999}
   ({\bf bottom}). 
Both images are filled contour plots with 7.5\arcmin resolution. The Galactic 
$b=0\degr$ plane is indicated by the dashed line. 
The CO emission from the central molecular disc has not been included (see Sec.\ref{sec:fux}), 
and the dust map excludes the area masked out in Fig.\ref{fig:ngp_dustlanes}. 
The contour represents $A_{Ks}=0.15$ mag.
Both the gas and dust are seen to be tilted with respect 
to the Galactic plane, with the negative longitude (far) side sitting above the plane and the positive 
longitude (near) side sitting below.}
   \label{fig:dustlanes}
\end{figure}

The sky distribution of the extinction due to the dust lanes is displayed on the top of 
Figure \ref{fig:dustlanes}. The quantity plotted is the integrated extinction owing to all the matter
 inside the ellipse of Fig. \ref{fig:ngp_dustlanes}. 
We choose sky projections to compare the extinction of the dust lanes with 
their CO emission to avoid any modelling required to deproject the raw CO data. 
In the bottom of Figure~\ref{fig:dustlanes} is the $l-b$
distribution of the gas as selected in Sec.\ref{sec:fux}.
The tilt  of the dust lanes relative to the Galactic plane is obvious in both the gas and the dust components,
with the negative longitude (far-side) part sitting above the plane and the positive longitude 
(near-side) part below the plane for both components. {\changes Also, both components are well 
contained within $|l|<10\degr$, showing that we are tracing a different structure to \cite{Calbet1996}.}

\begin{figure}
  \centering
    \includegraphics[trim = 3mm 0mm 0mm 3mm ,clip, width=0.8\linewidth]{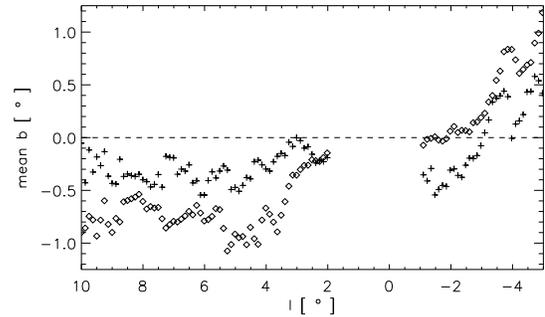}
  \caption{
    Position of the mean plane of dust (crosses) and molecular gas (diamonds) along the dust lanes in the Galactic bar. 
    The $b=0\degr$ plane is shown by the dashed line. The same large scale tilt is observed for both the gas and dust, however 
the dust appears closer to the $b=0\degr$ plane.
}
  \label{fig:meanplane}
\end{figure}

The mean plane of the dust/gas as a function of galactic longitude is given by
 $ \bar b(l) = {\sum_b b \cdot X(l,b)}/{\sum_b X(l,b)} $
where $X(l,b)$ is either the extinction due to dust 
or the intensity of the CO emission emanating from 
the dust lanes. The calculation of the mean plane for the dust uses lines of sight where $A_{Ks} \ge 0.15$ 
(indicated by the solid contour in the top image of Fig.\ref{fig:dustlanes}), in order to avoid low density lines 
of sight unrelated to the dust lanes.
The mean planes of the dust and the gas, calculated this way, 
are shown in Figure \ref{fig:meanplane}. Both the gas and the dust 
follow the same trend, however the dust is seen to be displaced from the gas.

The magnitude of the apparent displacement is larger at positive longitudes than at negative longitudes.  
The removal from the CO observations of the zone highlighted in Fig.\ref{A}b may bias the mean plane 
to more negative latitudes but, as the bulk of the dust lane CO emission arises from outside of this zone, the 
effect would be small. Likewise,
the contribution of the dust component outside the dust lanes has been minimised
using the high extinction lines of sight chosen using the 
ellipse of Sec.\ref{sec:marshall}. In fact, the difference in  apparent displacement  
is largely an effect of perspective, as the near side of the bar lies at positive longitudes.

These results suggest that, even though the large scale morphology of the dust lanes is similar for 
the two components studied here, there is an additional effect which causes their $l,b$ projections
 to be offset from one another. This could be due to an offset in $z$, a separation in the $x, y$ 
coordinates of a tilted plane or a combination of the two.

Offsets have been detected between dust and gas in external galaxies seen face on, where azimuthal offsets are easier
to discern than offsets in $z$.
For example, in M83 the dust {\it extinction} and CO emission in the eastern spiral arm are offset 
by as much as 700 pc \citep{Lord1991,Rand1999}.  
Three explanations are offered to explain the offset : 
1) The CO is heated by UV radiation from young stars; 
2) The CO is heated by low-energy cosmic rays; 
3) The components react differently to the spiral density wave.
In the case of the Milky Way bar, the first option is unlikely as 
star formation along the bar is likely hampered by the high shear there \citep{Rodriguez-Fernandez2006},
 and stars formed at the ends of the bar are not expected to migrate along it \citep{Cole2002}. 
Cosmic ray heating may be responsible for increased gas temperatures in the
 centre of the Galaxy \citep{Yusef-Zadeh2007} but these temperatures can 
also be explained by shock waves \citep{Lis2001}.
In the third case, the diffuse gas is trapped in a thin dust lane at the shock front and the molecular clouds 
form a broad ridge on the spiral arm. However this 
does not seem to correspond to our observations, where it is the dust that is seen to be
 more broadly distributed than 
the molecular gas (Fig.\ref{fig:dustlanes}).

Recently, \cite{Beck2005} observed the barred galaxy NGC 1097 in total and polarised intensity 
at $\lambda=$3.5 cm and  $\lambda=$6.2 cm. They 
found that around the bar the total intensity increases along the shearing shock, however 
the increase of the polarised emission occurs further downstream and is aligned with the bar's dust lanes.
They suggest that the large scale magnetic field is only weakly affected by the shock and that 
 the molecular clouds become decoupled from the magnetic field and are thus swept up by the bar's shearing shock.
The dust, coupled to the magnetic field, thus finds its way $\sim 400$ pc downstream from the CO.
Furthermore, 
they suggest that this may occur in most disc galaxies where large scale shocks trigger the formation or 
collapse of gas clouds. Our observations
 are consistent with this scenario, assuming that the gas and dust of the dust lanes orbit on a common 
and tilted plane.
 However, \cite{Beck2005} were not able to obtain direct observations of 
the molecular gas - dust separation in this galaxy.
In order to detect an offset of a few hundred parsecs between CO and dust
in an external galaxy would require high resolution and high sensitivity CO observations of a 
nearby barred galaxy with prominent dust lanes.

Instead of considering an offset on a common plane, it is possible that the observed offset is 
in effect a separation in $z$. As the gas and dust hit the shearing shock, it is possible that a fraction of 
this gas and dust fall towards the Galactic plane. Two possible mechanisms could be considered. Firstly, 
 \cite{Gomez2004} show using 3D MHD simulations that  
gas hitting a spiral arm shock can be deflected over the arm. 
Although there has been no study of this kind done on the bar shocks,  
there is also no evidence indicating that the gas and dust should be confined to the same plane. 
Secondly, the molecular gas and dust enter the shock front and merge with the dust lane at a given height above the Galactic plane. Due to the gravitational potential, the post-shock material eventually falls onto the Galactic plane creating a vertical distribution of azimuthally comoving gas and dust in the dust lane.

These mechanisms could explain a vertical distribution of gas and dust at the shock fronts but not 
a separation between them.
However, at the shock front the big grains of the dust (responsible for the NIR extinction)
are shattered \citep{Jones1994}, creating many smaller grains (responsible for UV extinction).
As the gas moves away from the 
shock front, the increased density speeds up the rate of coagulation between grains and, in particular,
 small grains on large ones.
If some of the gas and dust are heading towards the Galactic plane, 
the lack of CO at low $z$ could be explained by two processes: 
1) accretion of gaseous CO on to grains as an ice mantle, which is efficient for $A_{Ks}\ga0.35$ \citep{Draine2003}; 
2) dissociation of CO due to the increase in UV photon flux resulting from the decrease in the small grain abundance.
The dust layer would be found at systematically lower $z$ than the CO, as can be seen in 
Figs. \ref{fig:dustlanes} and \ref{fig:meanplane}.

\cite{Fux1999}  represented the tilts of the CO and  HI by a straight line with an angle of  
$\theta \approx 4.5\degr$ at their azimuth of $25\degr$, while mentioning that this representation is an 
oversimplification. 
The tilt of the dust component is not as easily represented by a straight line 
(Fig.\ref{fig:meanplane}). 
Its value is highly dependent on the segment of the dust lane which is fitted and the 
interpretation of its value depends on the scenario used to explain the offset discussed above. 

Finally, there are uncertainties in the dust position, namely in its distance along the line of sight. 
This uncertainty is estimated to be about 500 pc in the bar (Fig. \ref{fig:example}). 
Moreover the computation of the 3D extinction map made use of the Galactic model, 
hence it depends slightly on the parameters of this model \citep{Marshall2006}. 
This dependence might cause a systematic error of a few 100 pc in the distance of the dust along the line of
 sight, but not on its ($l,b$) position. It is expected that this error would 
affect the positive and negative longitudes similarly, 
having no influence on the tilt of the dust lanes in the ($l,b$) plane. 

\section{Conclusions}\label{sec:conclu}

We have identified the dominant dust lanes of the Galactic bar both by their near-infrared extinction and 
by their CO emission. The same large scale behaviour is observed in both components, that is 
a tilt going from $b<0\degr$ at positive $l$ to $b>0\degr$ at negative $l$. 
In the $l,b$ plane projections, the two components are seen to be displaced from one another.
This offset may be due to dust destruction at the shock front followed by reformation and/or ice mantle accretion
 downstream and at lower $z$, or
the molecular gas becoming decoupled from the magnetic field which resists 
the shearing shock of the Milky Way's bar and carries diffuse gas and dust further downstream on a tilted plane.
Further observations of our Galaxy and external galaxies, 
together with theoretical work are necessary in order to unambiguously interpret these observations.

Finally, the coincidence between the two independently recovered gas and dust distribution along the bar 
reinforces the interpretation from \cite{Fux1999} regarding the $l-V$ traces of the dust lanes 
in the observed HI and CO kinematics, and
it adds credibility to the Galactic model and the use
of it to determine the three dimensional extinction distribution.

\begin{acknowledgements}

The authors would like to thank G.Joncas, A.Jones, V.Guillet and A.Fletcher for helpful discussions.
D.J.Marshall was funded for part of this work by the Natural Sciences and Engineering Research Council of Canada
through its SRO programme.
This publication makes use of data products from the Two Micron All Sky Survey, which is a joint project of the 
University of Massachusetts and the Infrared Processing and Analysis Center/California Institute of Technology, 
funded by the National Aeronautics and Space Administration and the National Science Foundation. 
The CDSClient package 
  was used for the remote querying of the 2MASS dataset.
\end{acknowledgements}

\bibliographystyle{aa}
\bibliography{8967}

\begin{thebibliography}{24}
\expandafter\ifx\csname natexlab\endcsname\relax\def\natexlab#1{#1}\fi

\bibitem[{{Babusiaux} \& {Gilmore}(2005)}]{Babusiaux2005}
{Babusiaux}, C. \& {Gilmore}, G. 2005, \mnras, 358, 1309

\bibitem[{{Bania}(1977)}]{Bania1977}
{Bania}, T.~M. 1977, \apj, 216, 381

\bibitem[{{Beck} {et~al.}(2005){Beck}, {Fletcher}, {Shukurov}, {Snodin},
  {Sokoloff}, {Ehle}, {Moss}, \& {Shoutenkov}}]{Beck2005}
{Beck}, R., {Fletcher}, A., {Shukurov}, A., {et~al.} 2005, \aap, 444, 739

\bibitem[{{Calbet} {et~al.}(1996){Calbet}, {Mahoney}, {Hammersley}, {Garzon},
  \& {Lopez-Corredoira}}]{Calbet1996}
{Calbet}, X., {Mahoney}, T., {Hammersley}, P.~L., {Garzon}, F., \&
  {Lopez-Corredoira}, M. 1996, \apjl, 457, L27+

\bibitem[{{Cole} \& {Weinberg}(2002)}]{Cole2002}
{Cole}, A.~A. \& {Weinberg}, M.~D. 2002, \apjl, 574, L43

\bibitem[{{Dame} {et~al.}(2001){Dame}, {Hartmann}, \& {Thaddeus}}]{Dame2001}
{Dame}, T.~M., {Hartmann}, D., \& {Thaddeus}, P. 2001, \apj, 547, 792

\bibitem[{{de Vaucouleurs}(1964)}]{deVaucouleurs1964}
{de Vaucouleurs}, G. 1964, in IAU Symposium, Vol.~20, The Galaxy and the
  Magellanic Clouds, ed. F.~J. {Kerr}, 195--+

\bibitem[{{D\'esert} {et~al.}(1990){D\'esert}, {Boulanger}, \&
  {Puget}}]{Desert1990}
{D\'esert}, F.-X., {Boulanger}, F., \& {Puget}, J.~L. 1990, \aap, 237, 215

\bibitem[{{Draine}(2003)}]{Draine2003}
{Draine}, B.~T. 2003, \araa, 41, 241

\bibitem[{{Freudenreich}(1998)}]{Freudenreich1998}
{Freudenreich}, H.~T. 1998, \apj, 492, 495

\bibitem[{Fux(1999)}]{Fux1999}
Fux, R. 1999, \aap, 345, 787

\bibitem[{{Gerhard}(2002)}]{Gerhard2002}
{Gerhard}, O. 2002, in ASP Conf. Ser. 273, ed. G.~S. {Da Costa} \& H.~{Jerjen},
  73--+

\bibitem[{{G{\'o}mez} \& {Cox}(2004)}]{Gomez2004}
{G{\'o}mez}, G.~C. \& {Cox}, D.~P. 2004, \apj, 615, 744

\bibitem[{{Jones} {et~al.}(1994){Jones}, {Tielens}, {Hollenbach}, \&
  {McKee}}]{Jones1994}
{Jones}, A.~P., {Tielens}, A.~G.~G.~M., {Hollenbach}, D.~J., \& {McKee}, C.~F.
  1994, \apj, 433, 797

\bibitem[{{Lis} {et~al.}(2001){Lis}, {Serabyn}, {Zylka}, \& {Li}}]{Lis2001}
{Lis}, D.~C., {Serabyn}, E., {Zylka}, R., \& {Li}, Y. 2001, \apj, 550, 761

\bibitem[{{Liszt}(2006)}]{Liszt2006}
{Liszt}, H.~S. 2006, \aap, 447, 533

\bibitem[{{Lord} \& {Kenney}(1991)}]{Lord1991}
{Lord}, S.~D. \& {Kenney}, J.~D.~P. 1991, \apj, 381, 130

\bibitem[{{Marshall} {et~al.}(2006){Marshall}, {Robin}, {Reyl{\'e}},
  {Schultheis}, \& {Picaud}}]{Marshall2006}
{Marshall}, D.~J., {Robin}, A.~C., {Reyl{\'e}}, C., {Schultheis}, M., \&
  {Picaud}, S. 2006, \aap, 453, 635

\bibitem[{{Nagai} {et~al.}(2007){Nagai}, {Tanaka}, {Kamegai}, \&
  {Oka}}]{Nagai2007}
{Nagai}, M., {Tanaka}, K., {Kamegai}, K., \& {Oka}, T. 2007, \pasj, 59, 25

\bibitem[{{Rand} {et~al.}(1999){Rand}, {Lord}, \& {Higdon}}]{Rand1999}
{Rand}, R.~J., {Lord}, S.~D., \& {Higdon}, J.~L. 1999, \apj, 513, 720

\bibitem[{{Robin} {et~al.}(2003){Robin}, {Reyl{\'e}}, {Derri{\`e}re}, \&
  {Picaud}}]{Robin2003}
{Robin}, A.~C., {Reyl{\'e}}, C., {Derri{\`e}re}, S., \& {Picaud}, S. 2003,
  \aap, 409, 523

\bibitem[{Rodriguez-Fernandez {et~al.}(2006)Rodriguez-Fernandez, {Combes},
  {Martin-Pintado}, {Wilson}, \& {Apponi}}]{Rodriguez-Fernandez2006}
Rodriguez-Fernandez, N.~J., {Combes}, F., {Martin-Pintado}, J., {Wilson},
  T.~L., \& {Apponi}, A. 2006, \aap, 455, 963

\bibitem[{{Skrutskie} {et~al.}(2006){Skrutskie}, {Cutri}, {Stiening},
  {Weinberg}, {Schneider}, {Carpenter}, {Beichman}, {Capps}, {Chester},
  {Elias}, {Huchra}, {Liebert}, {Lonsdale}, {Monet}, {Price}, {Seitzer},
  {Jarrett}, {Kirkpatrick}, {Gizis}, {Howard}, {Evans}, {Fowler}, {Fullmer},
  {Hurt}, {Light}, {Kopan}, {Marsh}, {McCallon}, {Tam}, {Van Dyk}, \&
  {Wheelock}}]{Skrutskie2006}
{Skrutskie}, M.~F., {Cutri}, R.~M., {Stiening}, R., {et~al.} 2006, \aj, 131,
  1163

\bibitem[{{Yusef-Zadeh} {et~al.}(2007){Yusef-Zadeh}, {Wardle}, \&
  {Roy}}]{Yusef-Zadeh2007}
{Yusef-Zadeh}, F., {Wardle}, M., \& {Roy}, S. 2007, \apjl, 665, L123

\end{thebibliography}
\end{document}